\documentclass[
aps,
prd,
superscriptaddress,
nofootinbib,
floatfix]
{revtex4}
\usepackage{graphicx,
longtable,
color,
array,booktabs,
bm}

\begin{document}
\title{Majorana neutrino mass constraints\\
in the landscape of nuclear matrix elements}
%
\author{        	Eligio~Lisi}
\affiliation{   	Istituto Nazionale di Fisica Nucleare, Sezione di Bari, 
               		Via Orabona 4, 70126 Bari, Italy}
\author{        	Antonio~Marrone}
\affiliation{   	Dipartimento Interateneo di Fisica ``Michelangelo Merlin,'' 
               		Via Amendola 173, 70126 Bari, Italy}%
\affiliation{   	Istituto Nazionale di Fisica Nucleare, Sezione di Bari, 
               		Via Orabona 4, 70126 Bari, Italy}
\medskip
\begin{abstract}
We discuss up-to-date constraints on the Majorana neutrino mass $m_{\beta\beta}$ from neutrinoless \mbox{double} beta decay ($0\nu\beta\beta$) searches in experiments using different isotopes: KamLAND-Zen and EXO ($^{136}$Xe), GERDA and MAJORANA ($^{76}$Ge) and CUORE ($^{130}$Te). Best fits and upper bounds on $m_{\beta\beta}$ are explored in the general landscape of nuclear matrix elements (NME), as well as for specific NME values obtained in representative nuclear models. By approximating the likelihood of $0\nu\beta\beta$ signals through quadratic forms, the analysis of separate and combined isotope data becomes exceedingly simple, and allows to clarify various aspects of multi-isotope data combinations. In particular, we analyze the relative impact of different data in setting upper bounds on $m_{\beta\beta}$, as well as the conditions leading to nonzero $m_{\beta\beta}$ at best fit, for variable values of the NMEs. Detailed results on $m_{\beta\beta}$ from various combinations of data are reported in graphical and numerical form. Implications for future $0\nu\beta\beta$ data analyses and NME calculations are briefly discussed.
\end{abstract}
\medskip
\maketitle

\section{Introduction}
\label{Sec:Intro}

The process of neutrinoless double beta decay ($0\nu\beta\beta$),
\begin{equation}
(Z,\,A) \to (Z+2, \,A) + 2e^-\ ,
\label{0nubb}
\end{equation}
expected to occur for some candidate isotopes $(Z,\,A)$ if neutrinos are Majorana fermions, may be interpreted as a miniature event of leptonic matter creation or ``Little Bang,'' whose discovery would have profound implications for particle and nuclear physics and for cosmology \cite{Agostini:2022zub}. 

In the standard three-neutrino paradigm \cite{Zyla:2020zbs}, 
the process would be mediated by three Majorana neutrino mass states $\nu_i$ ($i=1,\,2,\,3$) mixed with the three known flavor states 
$\nu_\alpha$ ($\alpha=e,\,\mu,\,\tau$) via a unitary mixing matrix $U_{\alpha i}$, parametrized in terms of three mixing angles $(\theta_{12}\, \theta_{13}, \,\theta_{23})$, one (Dirac) phase $\delta$, and two (Majorana) phases $\varphi_{1,2}$. The relevant particle physics parameter is the effective Majorana neutrino mass $m_{\beta\beta}$, defined as
\begin{equation}
m_{\beta\beta} = |U^2_{e1} m_1 + U^2_{e2} m_2 + U^2_{e3} m_3|\ ,
\label{mbb}
\end{equation}
and related to the observable $0\nu\beta\beta$ decay half-life $T_i$ in each isotope $i=(Z,\,A)$ via
\begin{equation}
\frac{1}{T_i} = G_i M^2_i m^2_{\beta\beta}\ ,
\label{1/T}
\end{equation}
where $G_i$ is the phase space, and $M_i$ is the nuclear matrix element (NME) for the decay. 

\begin{figure}[t!]
\begin{minipage}[c]{0.58\textwidth}
\includegraphics[width=0.58\textwidth]{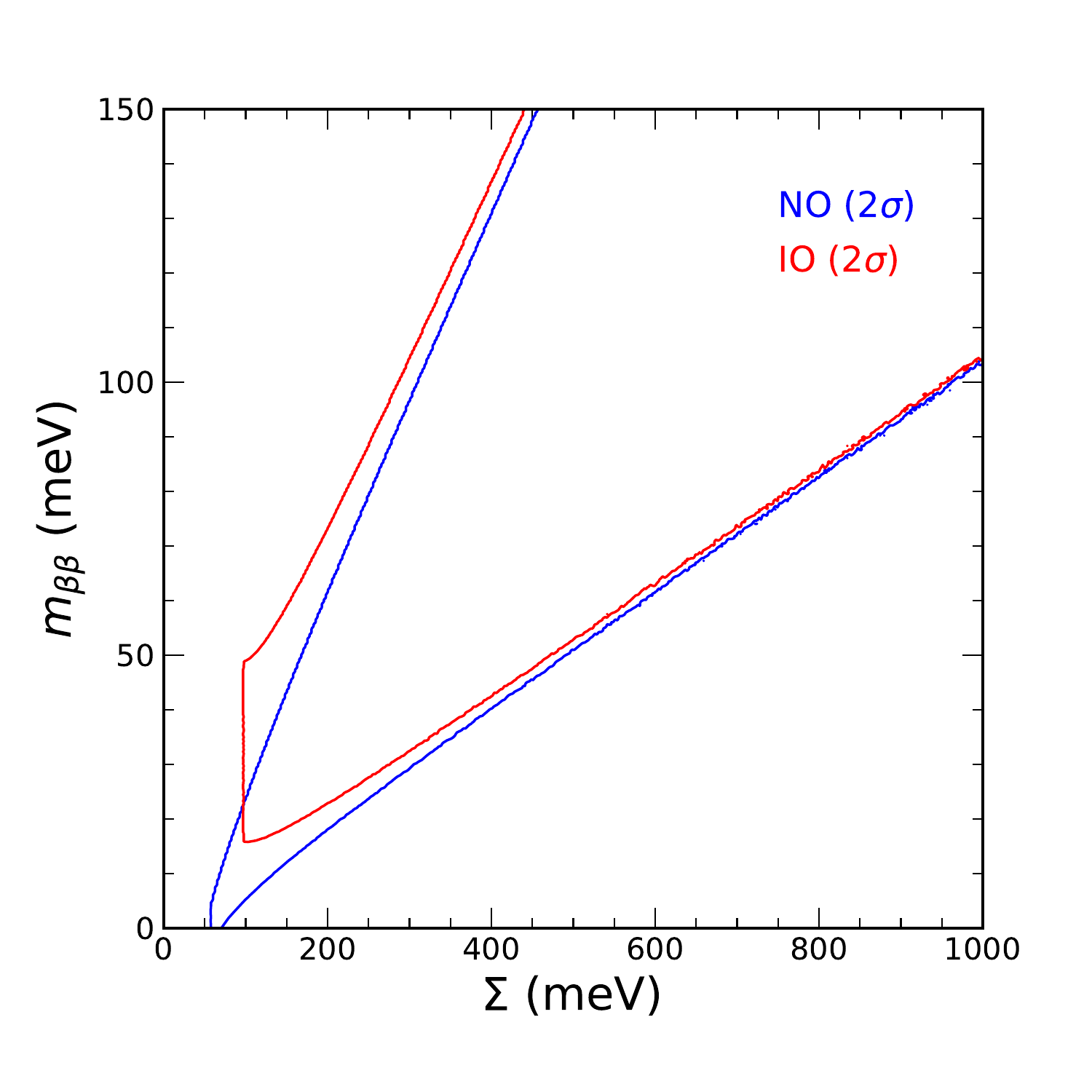}
\vspace*{-5mm}
\caption{\label{Fig_01}
\footnotesize Majorana $\nu$ mass $m_{\beta\beta}$ versus the sum of $\nu$ masses $\Sigma$ (both in units of $10^{-3}$~eV). 
\textcolor{black}{The points inside the blue (NO) and red (IO) wedge-shaped regions are allowed at $2\sigma$ by the global analysis of $\nu$ oscillation data.} 
Adapted from \protect\cite{Capozzi:2021fjo}.
} \end{minipage}
\end{figure}

It is useful to contrast the Majorana $\nu$ mass $m_{\beta\beta}$ with the sum of neutrino masses 
\begin{equation}
\Sigma = m_1+m_2+m_3\ , 
\end{equation}
that, being a source of gravity, can produce observable cosmological effects \cite{Zyla:2020zbs}. 
Figure~\ref{Fig_01} shows the regions  allowed in the $(\Sigma,\,m_{\beta\beta})$ plane at the $2\sigma$ level ($\Delta \chi^2=4$) by a global analysis of  neutrino oscillation data \cite{Capozzi:2021fjo}, for masses $m_i$ either in normal ordering (NO, $m_{1,2}<m_3$) or in inverted ordering (IO, $m_3<m_{1,2}$). For a given value of $\Sigma$, the vertical spread of $m_{\beta\beta}$ is mostly due to the unknown relative phases of the $U_{ei}$ matrix elements in Eq.~(\ref{mbb}).   

Current cosmological data provide typical upper bounds on $\Sigma$ at the level of $O(100)$~meV, that are more easily accommodated in NO than in IO \cite{Capozzi:2021fjo}; see also the overview of recent constraints on $\Sigma$ and their impact on $\nu$ mass ordering in \cite{Abazajian:2022ofy}. 
Several $0\nu\beta\beta$ decay searches are also exploring the O(100)~meV range for $m_{\beta\beta}$ \cite{Agostini:2022zub,Zyla:2020zbs}; in particular, the latest constraints from KamLAND-Zen  \cite{KamLAND-Zen:2022tow} 
($^{136}$Xe)
can plunge into the region $m_{\beta\beta}\sim \mathrm{few}\times 10$~meV for favorable values of the NME.%
\footnote{The range $m_{\beta\beta}\simeq 16$--49~meV, spanning the leftmost edge of the IO region in Fig.~\ref{Fig_01}, is often (but improperly) dubbed in $0\nu\beta\beta$ jargon as ``IO region,'' despite being compatible with both IO and NO (as well as the quasi-degenerate region at higher $m_{\beta\beta}$). The misnomer may originate from plots of $m_{\beta\beta}$ versus the lightest $\nu$ mass $m_l$, where two elongated stripes for IO and NO appear in log scale as $m_l \to 0$, see e.g.\ 
\cite{Agostini:2022zub,KamLAND-Zen:2022tow}. However, the asymptotic separation of such stripes has no physical relevance, since $m_l$ is not directly measurable and cannot be resolved with an accuracy better than an observable such as $\Sigma$. Projecting away $m_l$ (as in Fig.~\protect\ref{Fig_01}) makes the point clear.}
Other very sensitive $0\nu\beta\beta$ searches, all probing half-lives $T_i > 10^{25}$~y at 90\% C.L., have been performed by the experiments EXO \cite{Anton:2019wmi} ($^{136}$Xe), 
GERDA \cite{Agostini:2020xta} and MAJORANA \cite{Alvis:2019sil} ($^{76}$Ge), and CUORE \cite{Adams:2021rbc} ($^{130}$Te). 
\newpage

Building upon previous work \cite{Capozzi:2021fjo}, we discuss in detail how to combine current (Xe, Ge, Te) data for given NME values.%
\footnote{For simplicity,  we shall generally drop superscripts for the Xe, Ge, and Te isotopes.}
 The approach allows a do-it-yourself $0\nu\beta\beta$ global analysis in terms of $\chi^2$ functions with (up to) quadratic dependence on the signal strength $1/T_i$, which is a good approximation to recent results \cite{Capozzi:2021fjo,Caldwell:2017mqu,Biller:2021bqx}. In particular, we describe how to derive $m_{\beta\beta}$ constraints at a given confidence level, using both separate and combined (Xe, Ge, Te) data, for generic values of the nuclear matrix elements (the ``NME landscape''), as well as for representative NME values from different nuclear models. Our approach clarifies interesting aspects of the $0\nu\beta\beta$ data analysis, such as the relative importance of each isotope in determining (non)zero best fits and upper bounds for $m_{\beta\beta}$. 

The paper is structured as follows: In Section~\ref{Sec:Ingr} we describe the ingredients of our analysis in terms of notation, parametrization of experimental results for (Xe, Ge, Te), and associated NME's. In Section~\ref{Sec:Res} we discuss the main results of the analysis in terms of $m_{\beta\beta}$ constraints, by considering two qualitatively different situations: (1) cases where, a priori, $m_{\beta\beta}=0$ is preferred, and (2) more general cases where the best fit may be at $m_{\beta\beta}>0$. Upper bounds on $m_{\beta\beta}$ are explored both graphically and numerically in the NME landscape, by using separate and combined (Xe, Ge, Te) data. In Sec.~\ref{Sec:Sum} 
we summarize our results and comment on further applications and perspectives.

\section{Ingredients of the Analysis}
\label{Sec:Ingr}

In this Section we introduce the notation, the experimental results and their parametrization, the landscape of NME and
the phase space related to the
three isotopes Xe, Ge and Te.

\subsection{Notation and units}
\label{Subsec:Notation}

Following \cite{Capozzi:2021fjo}, we introduce the inverse half-life 
\begin{equation}
S_i = 1/T_i\ ,
\label{Sdef}
\end{equation}
that represents, up to a constant factor, the observable decay rate or signal strength in each $i=(Z,\,A)$ isotope.
Equation~(\ref{1/T}) reads then 
\begin{equation}
S_i = G_i M^2_i m^2_{\beta\beta}\ .
\label{Si}
\end{equation}
To keep the notation compact, we absorb in $G_i$ terms as  $1/m^{2}_e$ and $g^4_A$ (where $g_A=1.276$ 
\cite{Markisch:2018ndu} is the bare value of the axial-vector coupling), that are factorized out in other conventions.
\textcolor{black}{In particular, we can make contact with the notation of \cite{Agostini:2022zub}, where $1/T = G_{01} g_A^4M^2_{0\nu}m^2_{\beta\beta}/m^2_e$,
by identifying $G=G_{01}g_A^4/m^2_e$ and $M=M_{0\nu}$ for each isotope $i$. We also follow \cite{Agostini:2022zub} by taking   
the  $M_i$ as positive real numbers, referred to the
 bare value of $g_A$ (unless otherwise noticed). Qualitative effects of the so-called quenching of $g_A$ in nuclear matter \cite{Suhonen:2017krv} are separately commented below.}

Finally, the following units are adopted:
\begin{eqnarray}
\left[m_{\beta\beta}\right] &=& \mathrm{meV}\ ,\label{mbbunit}\\ 
\left[T_i\right] &=& 10^{26}\, y \ ,\label{Tunit}\\
\left[S_i\right] &=& 10^{-26}\, y^{-1}\ ,\label{Sunit}\\
\left[G_i\right] &=& 10^{-26}\, y^{-1}\, (\mathrm{meV})^{-2}\ .\label{Gunit}
\end{eqnarray}

\subsection{Experimental inputs and parametrizations}
\label{Subsec:ExpInput}

In principle, the $0\nu\beta\beta$ data analysis would be straightforward, if likelihood profiles were provided for the signal strength $S_i$ (or for $T_i$) in each experiment, e.g., in terms of a function $\chi^2_i(S_i)$. Barring error correlations among independent experiments, one should sum up the $\chi^2_i$ functions, express the $S_i$ in terms of $m_{\beta\beta}$ via Eq.~(\ref{Si}) for a given set of $M_i$, and map the resulting best fits and allowed regions for $m_{\beta\beta}$. In practice, experimental papers often focus on a single point of the likelihood profile (e.g., the $T_i$ bound at 90\% C.L.), whereas its shape has to be derived from supplementary information. 

A useful empirical fact, first noted in \cite{Caldwell:2017mqu} and further elaborated in  \cite{Capozzi:2021fjo}, is
that the functions $\chi^2_i(S_i)$ are often well approximated by (up to) quadratic forms in $S_i$; 
see also the recent results in \cite{Biller:2021bqx}. Such forms cover $0\nu\beta\beta$ decay searches
ranging from zero background 
(with poissonian, linear dependence on $S_i$) to large backgrounds (with gaussian, quadratic dependence on $S_i$) \cite{Capozzi:2021fjo}. In particular, we have checked that the quadratic approximation works very well also for the latest KamLAND-Zen data \cite{KamLAND-Zen:2022tow,Private}, up to $3\sigma$ level at least. 

Each experimental result is thus parametrized through a function $\Delta\chi^2_i(S_i)$ of the form:
\begin{equation}
\Delta \chi^2_i(S_i) = a_i S^2_i + b_i S_i + c_i\ ,
\label{Delta}
\end{equation}
where the offset $c_i$ is set by the condition that the minimum value $\Delta\chi^2_i=0$ is reached within the physical region 
$S_i\geq 0$, namely,
\begin{equation}
c_i = \left\{ \begin{array} {cl}
0 & \mathrm{for}~b_i\geq0\ , \\
b^2_i/4a_i & \mathrm{for}~b_i<0\ . \\
\end{array}\right.
\label{Offset}
\end{equation}
For $a_i>0$ the $\Delta\chi^2_i$ functions are parabolic, with a vertex placed at either $S_i=0$ (null result, $b_i=0$), 
or $S_i<0$ (negative fluctuation in the unphysical region, $b_i>0)$, or $S_i>0$  
(physical signal or positive fluctuation, $b_i<0$).
In the latter case, the offset $c_i$ guarantees $\Delta\chi^2_i=0$ at $S_i=-b_i/2a_i$. 
For $a_i=0$, the $\Delta\chi^2_i$ functions are linear. 
For the same isotope, the results of independent experiments are combined by 
summing their $\Delta \chi^2_i$'s, and readjusting the total offset as per Eq.~(\ref{Offset}).
In all cases, 90\% C.L.\ bounds on the half-life ($T_{90}=1/S_{90}$) 
are obtained by imposing $\Delta\chi^2_i(S_{90})=2.706$.  

Table~\ref{tab:abc}, updated from \cite{Capozzi:2021fjo} with the inclusion of the latest KamLAND-Zen results \cite{KamLAND-Zen:2022tow}, reports the coefficients of the parametrization in Eq.~(\ref{Delta}) and the $T_{90}$
bounds for the most sensitive current experiments ($T_{90}>0.1$), as well as for combinations of experiments using the same isotope.
For later purposes, we also consider hypothetical CUORE results for an exactly null signal, denoted as CUORE$^*$ (or Te$^*$). 

\begin{table}[b!]
\centering
\resizebox{.85\textwidth}{!}{\begin{minipage}{\textwidth}
\caption{\label{tab:abc} 
Coefficients of the quadratic parametrization  of $\Delta\chi^2_i$ in terms of the signal strength $S_i=1/T_i$. 
The first two columns report the isotope and the
names of the experiments or their combination. The next three columns report our evaluation of the 
coefficients $(a_i,\,b_i,\,c_i)$ for the various experiments (upper five rows) and for their combinations
in the same isotope (lower rows). The bottom row refers to the case of CUORE sensitivity for null result (tagged by $^*$).  
 The sixth column reports our 90\% C.L.\ 
($\Delta\chi^2=2.706$) half-life limits $T_{90}$, to be compared with the experimentally quoted one in the seventh column
(as taken from the reference in the eighth column, when applicabile).  
}
\begin{ruledtabular}
\begin{tabular}{rlrrrccc}
Isotope 	& Experiment or combination 		& $a_i~~~$ 	& $b_i~~~$ 	& $c_i~~~$ 	& $T_{90}/10^{26}\,\mathrm{y}$ & $T_{90}$ (expt.)
										& Reference \\ 
\hline
$^{136}$Xe	& KamLAND-Zen 				& 5.157		& 3.978		& 0.000		& 2.300 & 2.3
										& \cite{KamLAND-Zen:2022tow} \\
$^{136}$Xe	& EXO						& 0.440		& $-0.338$ 	& 0.065		& 0.350 & 0.35 
										& \cite{Anton:2019wmi}\\
$^{76}$Ge	& GERDA						& 0.000 	& \textcolor{black}{4.867} 	& 0.000 	& 1.800	& 1.8
										& \cite{Agostini:2020xta}\\
$^{76}$Ge	& MAJORANA					& 0.000		& \textcolor{black}{0.731} 	& 0.000		& 0.270 & 0.27 
										& \cite{Alvis:2019sil}\\
$^{130}$Te	& CUORE						& 0.245		& $-0.637$ 	& 0.414		& 0.216 & 0.22 
										& \cite{Adams:2021rbc}\\
\hline
$^{136}$Xe	& Xe  (KamLAND-Zen + EXO)			
										& 5.597		& 3.640		& 0.000		& 2.260 & \textemdash 
										& \textemdash \\
$^{76}$Ge	& Ge (GERDA + MAJORANA)		& 0.000		& 5.598		& 0.000		& 2.070 & \textemdash 
										& \textemdash  \\
$^{130}$Te	& Te (CUORE data as above)& 0.245		& $-0.637$ 	& 0.414		& 0.216 &  0.22 
										& \cite{Adams:2021rbc} \\
$^{130}$Te	& Te$^*$ (CUORE*, sensitivity)	& 0.245		& $0.000$ 	& 0.000		& 0.301 & 0.28 
										& \cite{Adams:2021rbc} \\
\end{tabular}
\end{ruledtabular}
\end{minipage}}
\end{table}

\newpage

\begin{figure}[t!]
\begin{minipage}[c]{0.74\textwidth}
\includegraphics[width=0.8\textwidth]{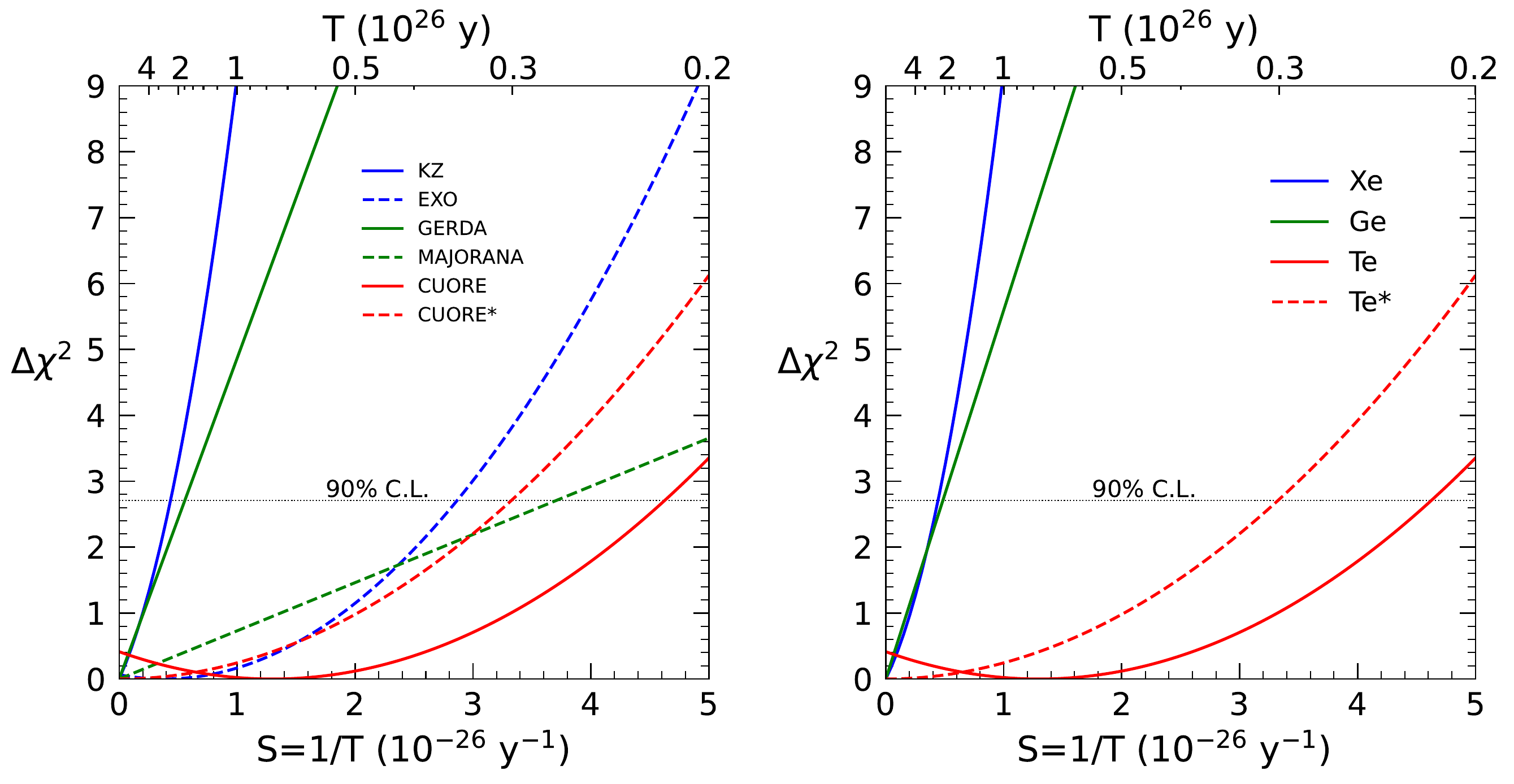}
\vspace*{-2mm}
\caption{\label{Fig_02}
\footnotesize $\Delta\chi^2$ functions in terms of the half-life $T$ (top abscissa) and of 
the signal strength $S = 1/T$ (bottom abscissa). Left and right panels: separate experiments and their combinations for the same isotope, respectively. Dotted horizontal lines intersect the curves at 90\% C.L. See the text for details.
} \end{minipage}
\end{figure}

Figure~\ref{Fig_02} shows the numerical information of Table~\ref{tab:abc} in graphical form; the left and right panels 
refer, respectively, to separate experiments and to same-isotope combinations (Xe, Ge, Te). 
A few remarks about these graphs and the numerics are in order. The case of linear $\Delta\chi^2$ functions applies to current GERDA and MAJORANA results, whose combination (denoted as Ge) sets
a bound $T_{90}=2.07$ stronger than for GERDA alone $(T_{90}=1.8)$. All the other experiments are characterized by parabolic functions.
KamLAND-Zen and EXO report, respectively, a negative and a positive 
fluctuation, that partly cancel in their combination (denoted as Xe). As a result, the Xe bound $T_{90}=2.26$ is
slightly weaker ($T_{90}=2.26$) than for KamLAND-Zen alone ($T_{90}=2.3$). Note that, for both the Ge and Xe combinations, it is $\Delta\chi^2_i=0$ at $S_i=0$. 

Results for the Te isotope depend on the single CUORE experiment, which shows a positive fluctuation
at the level of $0.64\sigma = [\Delta\chi^2_i(0)]^{1/2}$. As anticipated we consider, besides the real Te results, also hypothetical Te$^*$ results, where this fluctuation is canceled by setting $b_i=0$ (and thus also $c_i=0$). The half-time limit for Te$^*$ ($T_{90}=0.301$) is in 
reasonable agreement with the median sensitivity quoted by the CUORE experiment for null result ($T_{90}=0.28$). 
In Sec.~\ref{Sec:Res}, the 
combination of Xe, Ge, and Te$^*$ results (all with $\Delta\chi^2_i=0$ at $S_i=0$) will provide a simple
starting point, before discussing the full combination of Xe, Ge and Te constraints on $m_{\beta\beta}$. 

\vspace*{-2mm}
\subsection{Landscape of nuclear matrix elements}
\label{Subsec:Land}

In order to study the combination of $0\nu\beta\beta$ results in full generality, we consider 
unconstrained values of the nuclear matrix elements 
$M_\mathrm{Xe}$, $M_\mathrm{Ge}$, $M_\mathrm{Te}$ in the numerical range $M_i\in[0.2,\,20]$. Within this landscape,
we also consider representative $M_i$ values from four different approaches to nuclear modeling, including
the nuclear shell model (SM), the quasi-particle random phase approximation (QRPA), the energy-density functional theory (EDF),
and the interacting boson model (IBM). The NME values are taken from a recent compilation of results 
\cite{Menendez:2017fdf,Horoi:2015tkc,Coraggio:2020hwx,Mustonen:2013zu,Hyvarinen:2015bda,Simkovic:2018hiq,Fang:2018tui,Rodriguez:2010mn,LopezVaquero:2013yji,Song:2017ktj,Barea:2015kwa,Deppisch:2020ztt} as reported in \cite{Agostini:2022zub} 
\textcolor{black}{for the bare value of $g_A$}
(see Tab.~I therein), and are listed in Table~\ref{tab:NME} for the sake of completeness.  

\begin{figure}[h!]
\begin{minipage}[c]{0.85\textwidth}
\includegraphics[width=0.8\textwidth]{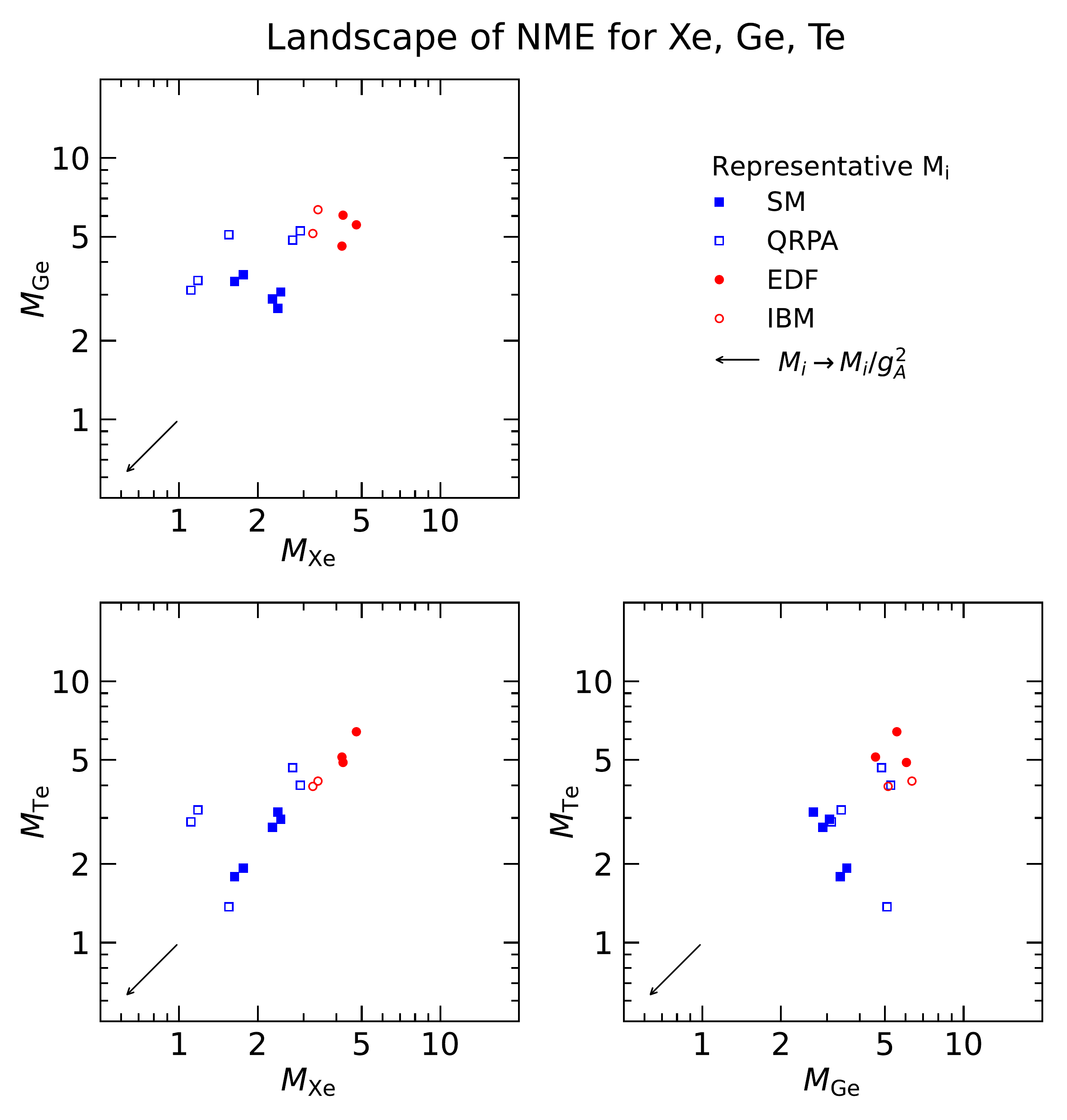}
\vspace*{-2mm}
\caption{\label{Fig_03}
\footnotesize Landscape of nuclear matrix elements $M_i=(M_\mathrm{Xe},\, M_\mathrm{Ge},\, M_\mathrm{Te})$,
in each of the three planes charted by pairs $(M_i,\,M_j)$.  Also shown are the representative $M_i$ values
reported in Table~\ref{tab:NME} from different models (SM, QRPA, EDF, IBM). The arrows show the effect
of rescaling each $M_i$ as $M_i/g^2_A$ (representing \textcolor{black}{typical} quenching effects). See the text for details.   
} \end{minipage}
\end{figure}

Figure~\ref{Fig_03} shows the NME landscape in each of the three planes charted by pairs $(M_i,\,M_j)$, together with
the representative $M_i$ values reported in Table~\ref{tab:NME}, which refer to the bare $g_A$. The issue of the 
effective $g_A$ value to be used in nuclear matter, 
\textcolor{black}{either bare or quenched by a factor $q$ ($g_A\to q\, g_A$ with $q<1$)},
is largely debated 
\textcolor{black}{and model-dependent}
\cite{Agostini:2022zub,Suhonen:2017krv}.
\textcolor{black}{For NMEs dominated by the axial-vector (Gamow-Teller) component (as it is often the case), the leading quenching effect would amount to rescaling the product $G_i M_i^2$ by a factor $q^4 \sim g_A^4$, that can be assumed to operate on $M_i$ only ($M_i\to q^2 M_i$) if $G_i$ is kept constant.}

\textcolor{black}{As a representative quenching effect one may consider, e.g., the typical case $q\, g_A \simeq 1$, namely, $q\sim 1/g_A$, leading to an approximate rescaling}   
$M_i\to M_i/g_A^2$, as shown in each panel of Fig.~\ref{Fig_03} by an arrow (applicable to any marked point). Stronger quenching would be associated to longer arrows (not shown). 
\textcolor{black}{On the other hand, quenching effects would be weaker for sizeable NME vector (Fermi) components, not scaling
with $q^2$. Moreover, some $M_i$ calculations
may exhibit a milder dependence on $q g_A$ for different reasons.} In some QRPA calculations, e.g.,
$2\nu\beta\beta$ data are used to adjust the 
particle-particle parameter $g_{pp}$, partly trading the effect of quenching $g_A$ from its bare value to unity \cite{Faessler:2008xj}. 
In the same approach, large statistical covariances are observed among the $M_i$ values 
for different isotopes, inducing noticeable effects on $m_{\beta\beta}$ constraints  
\cite{Faessler:2008xj}, as recently discussed in \cite{Capozzi:2021fjo}. 
The marked points in Fig.~\ref{Fig_03} also seem to suggest
an overall positive correlation (possibly enhanced by quenching effects) but,
since they do not represent a statistical distribution, their covariances (if any) will be ignored.

\begin{table}[t!]
\centering
\resizebox{.35\textwidth}{!}
{\begin{minipage}{0.45\textwidth}
\caption{\label{tab:NME} 
Representative nuclear matrix elements $(M_\mathrm{Xe},\, M_\mathrm{Ge},\, M_\mathrm{Te})$ computed within four
different models (SM, QRPA, EDF, and IMB) for bare value of $g_A$. Adapted from \cite{Agostini:2022zub}. 
}
\begin{ruledtabular}
\begin{tabular}{cccccc}
 & $M_\mathrm{Xe}$ & $M_\mathrm{Ge}$ & $M_\mathrm{Te}$ & Reference & Model \\ 
\hline
1 & 2.28 & 2.89 & 2.76 & \cite{Menendez:2017fdf} &     \\
2 & 2.45 & 3.07 & 2.96 & \cite{Menendez:2017fdf} &     \\
3 & 1.63 & 3.37 & 1.79 & \cite{Horoi:2015tkc}    &  SM \\
4 & 1.76 & 3.57 & 1.93 & \cite{Horoi:2015tkc}    &     \\
5 & 2.39 & 2.66 & 3.16 & \cite{Coraggio:2020hwx} &     \\
\hline
6 & 1.55 & 5.09 & 1.37 & \cite{Mustonen:2013zu}  &     \\
7 & 2.91 & 5.26 & 4.00 & \cite{Hyvarinen:2015bda}&     \\
8 & 2.72 & 4.85 & 4.67 & \cite{Simkovic:2018hiq} & QRPA\\
9 & 1.11 & 3.12 & 2.90 & \cite{Fang:2018tui}     &     \\
10& 1.18 & 3.40 & 3.22 & \cite{Fang:2018tui}     &     \\
\hline
11& 4.20 & 4.60 & 5.13 & \cite{Rodriguez:2010mn} &     \\
12& 4.77 & 5.55 & 6.41 & \cite{LopezVaquero:2013yji}& EDF \\
13& 4.24 & 6.04 & 4.89 & \cite{Song:2017ktj}     &     \\
\hline
14& 3.25 & 5.14 & 3.96 & \cite{Barea:2015kwa}    &     \\
15& 3.40 & 6.34 & 4.15 & \cite{Deppisch:2020ztt} & IBM     
\end{tabular}
\end{ruledtabular}
\end{minipage}}
\end{table}

\textcolor{black}{Summarizing, the NME landscape in Fig.~\ref{Fig_03} is meant to cover a wide and continuous range of 
$M_i$ values, either unquenched or arbitrarily quenched. Graphical results will be shown for unconstrained $M_i$ values in this landscape.
Marked points in Fig.~\ref{Fig_03} are meant to represent typical 
unquenched $M_i$ (central values) as taken from the literature (see Table~\ref{tab:NME}), while the arrows provide visual guidance for
typical quenching effects ($q\,g_A\sim 1$).}

 \subsection{Phase space}
\label{Subsec:Phase}

The last ingredient is represented by the phase space $G_i$ for $0\nu\beta\beta$ decay in $i=\mathrm{Xe}$, Ge and Te, that we take 
from the calculation in \cite{Deppisch:2020ztt}. In our notation and units:
\begin{eqnarray}
G_\mathrm{Xe} &=& 14.78 \times 10^{-6}\ , \\
G_\mathrm{Ge} &=& \phantom{0}2.40 \times 10^{-6}\ ,\\
G_\mathrm{Te} &=& 14.42 \times 10^{-6}\ .
\end{eqnarray}
 
\textcolor{black}{The phase space uncertainties \cite{Stoica:2019ajg} are much smaller than those related to $0\nu\beta\beta$ data and are not considered herein.} 
 
\vspace*{-1mm} 
\section{Constraints on the Majorana neutrino mass}
\label{Sec:Res}

The previously discussed functions $\Delta\chi^2_i(S_i)$ can be recast in terms of quadratic
functions of $m^2_{\beta\beta}$ through Eq.~(\ref{Si}):
\begin{equation}
\Delta \chi^2_i=\alpha_i m^4_{\beta\beta} + \beta_i m^2_{\beta\beta} + \gamma_i\ ,
\label{Deltambb}
\end{equation}
where  the offset $\gamma_i$ is set by
\begin{equation}
\gamma_i = \left\{ \begin{array} {cl}
0 & \mathrm{for}~\beta_i\geq0\ , \\
\beta^2_i/4\alpha_i & \mathrm{for}~\beta_i<0\ . \\
\end{array}\right.
\label{Offset2}
\end{equation}
The best-fit value of $m_{\beta\beta}$ is set by:
\begin{equation}
\Delta\chi_i^2=0 \to
m_{\beta\beta} = \left\{ \begin{array} {cl}
0 & \mathrm{for}~\beta_i\geq0\ , \\
(-\beta_i/2\alpha_i)^{1/2}  & \mathrm{for}~\beta_i<0\ . \\
\end{array}\right.
\label{BestFit}
\end{equation}

Table~\ref{tab:abcgreek} reports the parametric coefficients $(\alpha_i,\,\beta_i,\,\gamma_i)$ for the Xe, Ge, Te and Te* cases,
with their explicit dependence on the matrix elements $M_\mathrm{Xe}$, $M_\mathrm{Ge}$, $M_\mathrm{Te}$.

\begin{table}[b!]
\centering
\resizebox{.5\textwidth}{!}{\begin{minipage}{0.5\textwidth}
\caption{\label{tab:abcgreek} 
Coefficients of the quadratic parametrization $\Delta\chi^2_i=\alpha_i m^4_{\beta\beta} + \beta_i m^2_{\beta\beta} + \gamma_i$
for the cases Xe, Ge, Te and Te$^*$. 
}
\begin{ruledtabular}
\begin{tabular}{l ccc}
Case & $\alpha_i$ & $\beta_i$ & $\gamma_i$ \\	
\hline
Xe		& $1.223\times 10^{-9}  M^4_\mathrm{Xe}$	& $\phantom{+}5.380\times 10^{-5}M^2_\mathrm{Xe}$ 	& 0 \\
Ge		& 0											& $\phantom{+}1.344\times 10^{-5}M^2_\mathrm{Ge}$ 	& 0 \\
Te		& $5.094\times 10^{-11} M^4_\mathrm{Te}$	& $-9.186\times10^{-6}M^2_\mathrm{Te}$  & 0.414	 \\
Te$^*$	& $5.094\times 10^{-11} M^4_\mathrm{Te}$	& 0 									& 0 
\end{tabular}
\end{ruledtabular}
\end{minipage}}
\end{table}

Constraints on $m_{\beta\beta}$ from two or more isotopes are obtained by summing the corresponding $\Delta\chi^2_i$, and by adjusting the offset so that it obeys Eq.~(\ref{Offset2}), namely:
$\Delta \chi^2 = \alpha m^4_{\beta\beta} + \beta m^2_{\beta\beta} + \gamma $,
where $\alpha=\sum_i\alpha_i$, $\beta=\sum_i\beta_i$, and $\gamma =0$  for $\beta\geq 0$ ($\gamma = \beta^2/4\alpha$ otherwise).
Bounds on $m_{\beta\beta}$ at a given confidence level are obtained  by solving  
\begin{equation}
\Delta\chi^2 (m_{\beta\beta})=\Delta_\mathrm{CL}\ ,
\label{CL}
\end{equation}
where, e.g., $\Delta_\mathrm{CL}=2.706$, 4 and 9 for limits at 90\% C.L., $2\sigma$ and $3\sigma$, respectively.

Two qualitatively different cases arise from current results: 
(a) for Xe, Ge and Te$^*$, either separately or in combination, there is no offset $\gamma$ and the $\Delta\chi^2$ function is zeroed at 
$m_{\beta\beta}=0$; (b) for Te results, characterized by $\gamma_i>0$ (positive fluctuation), 
the combination with Xe or Ge (or both) may lead to $\gamma>0$, implying a nonzero
Majorana neutrino mass at best fit: $m_{\beta\beta}=[-\beta/2\alpha]^{1/2}$. We discuss separately these two cases below.

\vspace*{-2mm} 
\subsection{Combination of Xe, Ge, Te$^*$ constraints}
\label{sec:Te*}

In this section we consider current constraints from Xe, Ge, and from the pseudo-experiment Te$^*$ (corresponding to null signal in CUORE). In this case the analysis is straightforward, since the best fit is $m_{\beta\beta}=0$ for all isotopes and
their combinations (independently of the NME). The analysis including real Te data will bring forward cases with $m_{\beta\beta}>0$ at best fit, as discussed in the next Section.
 
We consider both separate and combined Xe, Ge and Te$^*$ constraints, as obtained by summing up the corresponding
functions defined in Table~\ref{tab:abcgreek}, $\Delta\chi^2=\sum_i\Delta\chi^2_i=\alpha m^4_{\beta\beta} + \beta m^2_{\beta\beta}$. 
For a given choice of $M_i$, 
upper limits on $m_{\beta\beta}$ are obtained by solving 
$\Delta_\mathrm{CL}=\alpha m^4_{\beta\beta} + \beta m^2_{\beta\beta}$. For definiteness we set $\Delta_\mathrm{CL}=4$ ($2\sigma$ bounds), unless otherwise specified. 

Table~\ref{tab:Te*} reports the upper bounds on $m_{\beta\beta}$, for each of the representative ($M_\mathrm{Xe},\,M_\mathrm{Ge},\,M_\mathrm{Te}$) calculations listed in Table~\ref{tab:NME}. Concerning constraints from single isotopes,
in most cases Xe sets the strongest bound, followed by weaker ones from Ge and Te$^*$. 
However, for the cases numbered as 9 and 10 (QRPA), the bounds from Xe, Ge and Te$^*$ are comparable to each other, and
for case 6 (QRPA) the Ge bound actually prevails over the Xe (and Te$^*$) bound. Notice that
such a hierarchy of $m_{\beta\beta}$ constraints may change at different confidence levels, 
since the $S_i$ bounds scale up at different rates (see Fig.~\ref{Fig_02}).
In Table~\ref{tab:Te*} the
combination of pairs of constraints  improves appreciably upon each separate constraint; 
the relative balance in each pair is highlighted below. 
Finally, the total combination Xe+Ge+Te$^*$ provides even stronger bounds on $m_{\beta\beta}$, that
range from a  minimum of 38.5 meV (case 12, EDF) to a maximum of 120.4 meV (case 9, QRPA) at $2\sigma$. 

\begin{table}[t!]
\centering
\resizebox{.84\textwidth}{!}
{\begin{minipage}{1.0\textwidth}
\caption{\label{tab:Te*} 
Bounds on $m_{\beta\beta}/$meV at $2\sigma$ level from Xe, Ge and Te$^*$ results, both
separately and in combination, for each of the 15 representative NME calculations listed in Tab.~\ref{tab:NME}.
The best-fit value is $m_{\beta\beta}=0$ in all cases.
}
\begin{ruledtabular}
\begin{tabular}{lrrrrrrrrrrrrrrr}
  			& 1 & 2 & 3 & 4 & 5 & 6 & 7 & 8 & 9 & 10 & 11 & \phantom{12.}12 & 13 & 14 & 15 \\  	
\hline
Xe			&  86.9&  80.9& 121.6& 112.6&  82.9& 127.9&  68.1&  72.9& 178.6& 168.0&  47.2&  41.6&  46.7&  61.0&  58.3\\
Ge			& 188.8& 177.7& 161.9& 152.8& 205.1& 107.2& 103.7& 112.5& 174.9& 160.5& 118.6&  98.3&  90.3& 106.2&  86.1\\
Te$^*$		& 191.8& 178.8& 295.7& 274.3& 167.5& 386.4& 132.3& 113.4& 182.5& 164.4& 103.2&  82.6& 108.3& 133.7& 127.6\\
Xe+Ge		&  81.0&  75.5& 101.9&  95.0&  78.5&  86.8&  59.3&  63.7& 132.0& 122.7&  44.8&  39.2&  42.8&  54.8&  50.4\\
Xe+Te$^*$	&  85.7&  79.8& 120.4& 111.6&  81.3& 127.4&  66.6&  69.2& 147.2& 135.3&  46.5&  40.7&  46.2&  60.1&  57.5\\
Ge+Te$^*$	& 149.7& 140.2& 155.6& 146.5& 142.2& 106.9&  91.3&  88.8& 140.7& 127.8&  85.8&  69.5&  77.5&  92.9&  79.4\\
Xe+Ge+Te$^*$&  80.1&  74.7& 101.3&  94.4&  77.2&  86.7&  58.4&  61.5& 120.4& 110.7&  44.2&  38.5&  42.4&  54.2&  49.9\\	 
\end{tabular}
\end{ruledtabular}
\vspace*{-2mm}
\end{minipage}}
\end{table}

\begin{figure}[b!]
\vspace*{-3mm}
\begin{minipage}[c]{0.85\textwidth}
\includegraphics[width=0.75\textwidth]{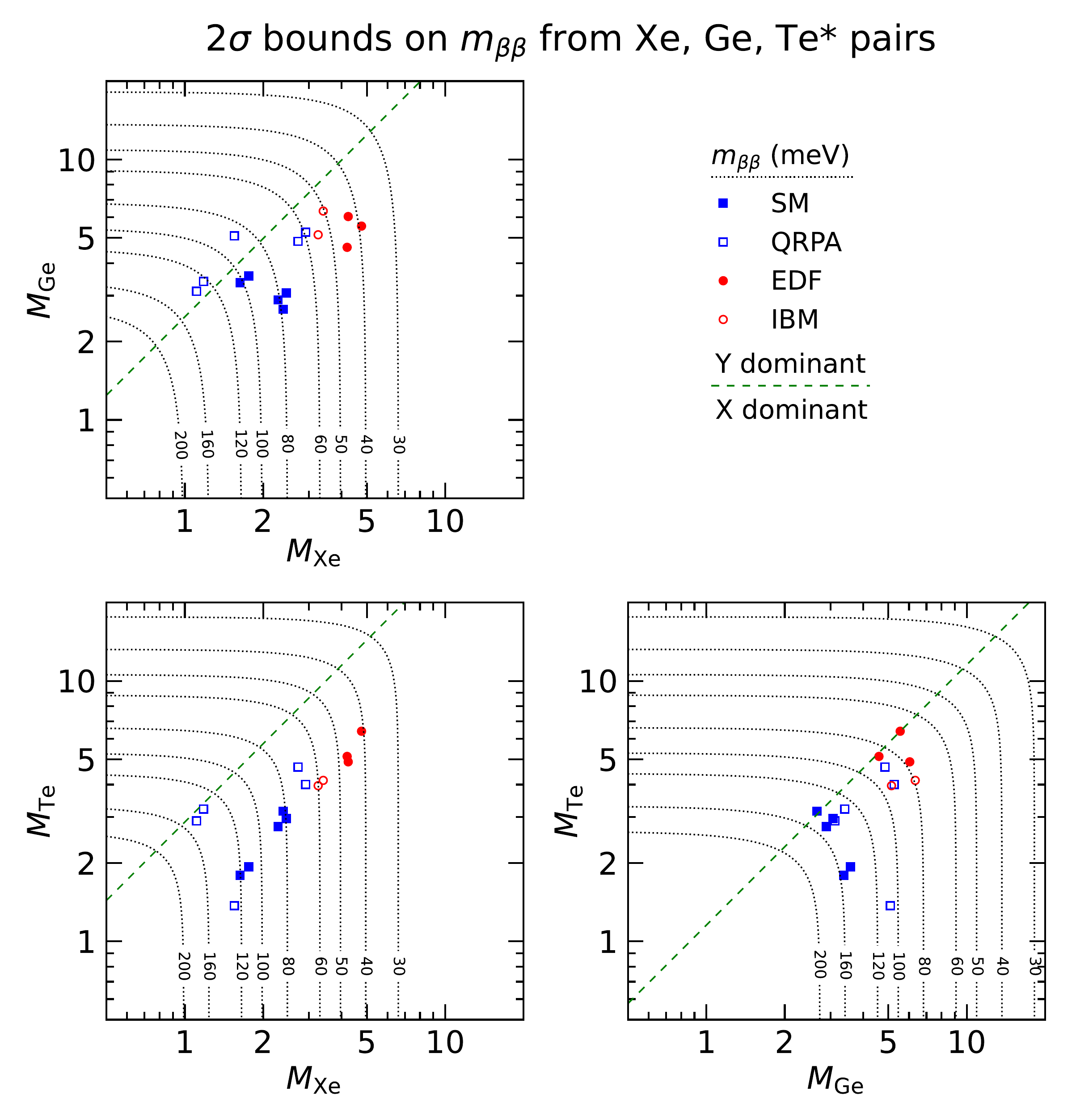}
\vspace*{-2mm}
\caption{\label{Fig_04}
\footnotesize Isolines of $m_{\beta\beta}$ bounds (at $2\sigma$ level) in the landscape of nuclear matrix elements, obtained from the combination of any
two results among Xe, Ge and Te$^*$. In each panel, the bounds are dominated by the isotope on the $y$ ($x$) axis, 
in the region above (below) the  dashed line. See the text for details.
} \end{minipage}
\end{figure}

Figure~\ref{Fig_04} shows isolines of the $2\sigma$ bounds on $m_{\beta\beta}$, 
as derived by combining any two pairs among Xe, Ge and Te$^*$, for unconstrained
values of the NME. 
In each panel, single-isotope bounds are asimptotically 
recovered along each axis, for vanishing matrix element on the other axis. 
When both matrix elements are sizeable, the joint bound improves upon separate ones. In
 particular, at each marked point, the bounds in Table~\ref{tab:Te*} are recovered 
for the corresponding NME and pair of isotopes.

For each $(M_x,\,M_y)$ panel and $(x,\,y)$ 
isotope pair, the condition for the dominance of one isotope constraint over the other is easily derived.
The two isotopes contribute equally to $\Delta_\mathrm{CL}$ when $\Delta \chi^2_x = \Delta_\mathrm{CL}/2=\Delta \chi^2_y$. 
The solutions to these equations read $M_x m_{\beta\beta}=\xi_x$ and $M_y m_{\beta\beta}=\xi_y$, where $\xi_{x,y}$ are positive numbers. For $\Delta_\mathrm{CL}=4$ it is
$M_\mathrm{Te}/M_\mathrm{Ge}=1.154$, $M_\mathrm{Ge}/M_\mathrm{Xe}=2.488$, and $M_\mathrm{Te}/M_\mathrm{Xe}=2.871$,
 shown as a dashed line in each panel of Fig.~\ref{Fig_04}. Along the dashed line, the two isotopes  
contribute with equal strength to the $2\sigma$ upper bound; above the dashed line,
the $y$-axis isotope dominates over the $x$-axis one, and viceversa. In this way one gets a graphical
interpretation of the hierarchy of bounds for different NME, that was
inferred from numerical inspection of Table~\ref{tab:Te*}.

\subsection{Combination of Xe, Ge, Te constraints}
\label{sec:Te}

In this section we consider real Te data, as opposed to the previous cases including Te* pseudo-data.  
The slight preference of Te data from CUORE for a nonzero signal (as compared with Xe, Ge and Te$^*$,  
see Fig.~\ref{Fig_02}) brings forward new features of multi-isotope data constraints, although still at embryonic stages.

In general one may expect that, for relatively small values of $M_\mathrm{Te}$ (with respect to $M_\mathrm{Xe}$ and
$M_\mathrm{Ge}$), the Xe+Ge results will dominate
over Te, keeping the best fit at $m_{\beta\beta}=0$. However, for increasing $M_\mathrm{Te}$, the Te
results will eventually prevail and set $m_{\beta\beta}>0$ at $\Delta\chi^2=0$, affecting also upper bounds at some value 
$\Delta_\mathrm{CL}$. 

This situation anticipates what could happen 
with future and more accurate $0\nu\beta\beta$ data: their combination may (or may not) be 
consistent with some indications for nonzero $m_{\beta\beta}$, depending on both the data and the 
NME values. A future preference for $m_{\beta\beta}>0$ might even lead to lower bounds on 
$m_{\beta\beta}$, either separately or in combination, depending in part on (un)favorable values
of the NME. Eventually, precise multi-isotope data might even test specific NME's by selecting allowed ratios $M_x/M_y$  \cite{Bilenky:2002ga,Bilenky:2004um} namely,
slanted allowed stripes in the NME landscape of Fig.~\ref{Fig_03}.%
\footnote{An overall NME rescaling factor $\lambda$ 
($M_{x,y}\to\lambda M_{x,y} $) is degenerate with an inverse rescaling of 
the Majorana mass ($m_{\beta\beta}\to m_{\beta\beta}/\lambda$). }

In our approach, the occurrence of $m_{\beta\beta}>0$ at best fit is simply signaled, for a single isotope, 
by a coefficient $\beta_i<0$ in the $\Delta\chi^2$ function (currently occurring only for Te) and, for any combination 
of multi-isotope data, by a negative coefficient $\beta=\sum_i\beta_i<0$.  The best-fit value of $m_{\beta\beta}$ is then
$m_{\beta\beta}=(-\beta/2\alpha)^{1/2}$ with $\alpha=\sum_i{\alpha_i}$, and its specific value depends on the NME's. 
For $\beta<0$, the offset must be taken as $\gamma=\beta^2/4\alpha$. In all cases, upper bounds at a
chosen confidence levels
are set by $\Delta\chi^2=\Delta_\mathrm{CL}$. Table~\ref{tab:Te} shows the numerical results from current Xe, Ge and Te data, regarding the $2\sigma$ limits (lower part)
and the best-fit values (lower half) of $m_{\beta\beta}$. In the upper half, the rows corresponding the
 Xe, Ge and Ge+Xe are unchanged with respect to Table~\ref{tab:Te*}, but are repeated
for completeness.

\begin{table}[b!]
\centering
\resizebox{.84\textwidth}{!}
{\begin{minipage}{1.0\textwidth}
\caption{\label{tab:Te} 
Upper half: Bounds on $m_{\beta\beta}/$meV at $2\sigma$ level from Xe, Ge and Te results, both
separately and in combination, for each of the 15 representative NME calculations listed in Tab.~\ref{tab:NME}. 
Lower half: Corresponding best-fit values of $m_{\beta\beta}$. 
}
\begin{ruledtabular}
\begin{tabular}{lrrrrrrrrrrrrrrr}
  			& 1 & 2 & 3 & 4 & 5 & 6 & 7 & 8 & 9 & 10 & 11 & \phantom{12.}12 & 13 & 14 & 15 \\  	
\hline
Xe			&  86.9&  80.9& 121.6& 112.6&  82.9& 127.9&  68.1&  72.9& 178.6& 168.0&  47.2&  41.6&  46.7&  61.0&  58.3\\
Ge			& 188.8& 177.7& 161.9& 152.8& 205.1& 107.2& 103.7& 112.5& 174.9& 160.5& 118.6&  98.3&  90.3& 106.2&  86.1\\
Te			& 220.5& 205.6& 340.0& 315.3& 192.6& 444.2& 152.1& 130.3& 209.9& 189.0& 118.6&  94.9& 124.5& 153.7& 146.6\\
Xe+Ge		&  81.0&  75.5& 101.9&  95.0&  78.5&  86.8&  59.3&  63.7& 132.0& 122.7&  44.8&  39.2&  42.8&  54.8&  50.4\\
Xe+Te		&  89.6&  83.4& 124.9& 115.7&  85.6& 130.4&  70.4&  75.0& 167.5& 154.5&  48.6&  42.9&  48.1&  62.9&  60.1\\
Ge+Te		& 174.1& 163.2& 170.1& 160.5& 166.5& 109.6& 104.4& 103.3& 163.4& 148.6& 100.3&  81.3&  89.2& 106.5&  89.2\\
Xe+Ge+Te	&  83.5&  77.9& 104.5&  97.4&  81.2&  88.0&  61.4&  66.2& 135.0& 124.7&  46.2&  40.5&  44.1&  56.5&  51.9\\
\hline
Xe			&     0&     0&     0&     0&     0&     0&     0&     0&     0&     0&     0&     0&     0&     0&     0\\
Ge			&     0&     0&     0&     0&     0&     0&     0&     0&     0&     0&     0&     0&     0&     0&     0\\
Te			& 108.8& 101.4& 167.7& 155.6&  95.0& 219.2&  75.1&  64.3& 103.5&  93.2&  58.5&  46.8&  61.4&  75.8&  72.4\\
Xe+Ge		&     0&     0&     0&     0&     0&     0&     0&     0&     0&     0&     0&     0&     0&     0&     0\\
Xe+Te		&     0&     0&     0&     0&     0&     0&     0&     0&  31.7&  36.0&     0&     0&     0&     0&     0\\
Ge+Te		&     0&     0&     0&     0&     0&     0&     0&     0&     0&     0&     0&     0&     0&     0&     0\\
Xe+Ge+Te	&     0&     0&     0&     0&     0&     0&     0&     0&     0&     0&     0&     0&     0&     0&     0\\
\end{tabular}
\end{ruledtabular}
\end{minipage}}
\end{table}

Let us first comment on the $m_{\beta\beta}$ best fits in the lower half of Table~\ref{tab:Te}.
Of course, the Te row displays nonzero results, with $m_{\beta\beta}$ scaling as $1/M_\mathrm{Te}$.
In almost all cases including Te (combined with Xe or Ge or both), the positive contributions to
$\beta=\sum_i\beta_i$ from $i=$~Xe and
Ge are never erased by the negative contribution from Te, and 
$m_{\beta\beta}=0$ is preferred. Only for cases 9 and 10 (QRPA), 
it turns out that the large ratio $M_\mathrm{Te}\simeq 2.6 M_\mathrm{Xe}$ 
makes Te prevail over Xe in the corresponding Xe+Te combinations, which show nonzero best fits.

\begin{figure}[t!]
\begin{minipage}[c]{0.85\textwidth}
\includegraphics[width=0.73\textwidth]{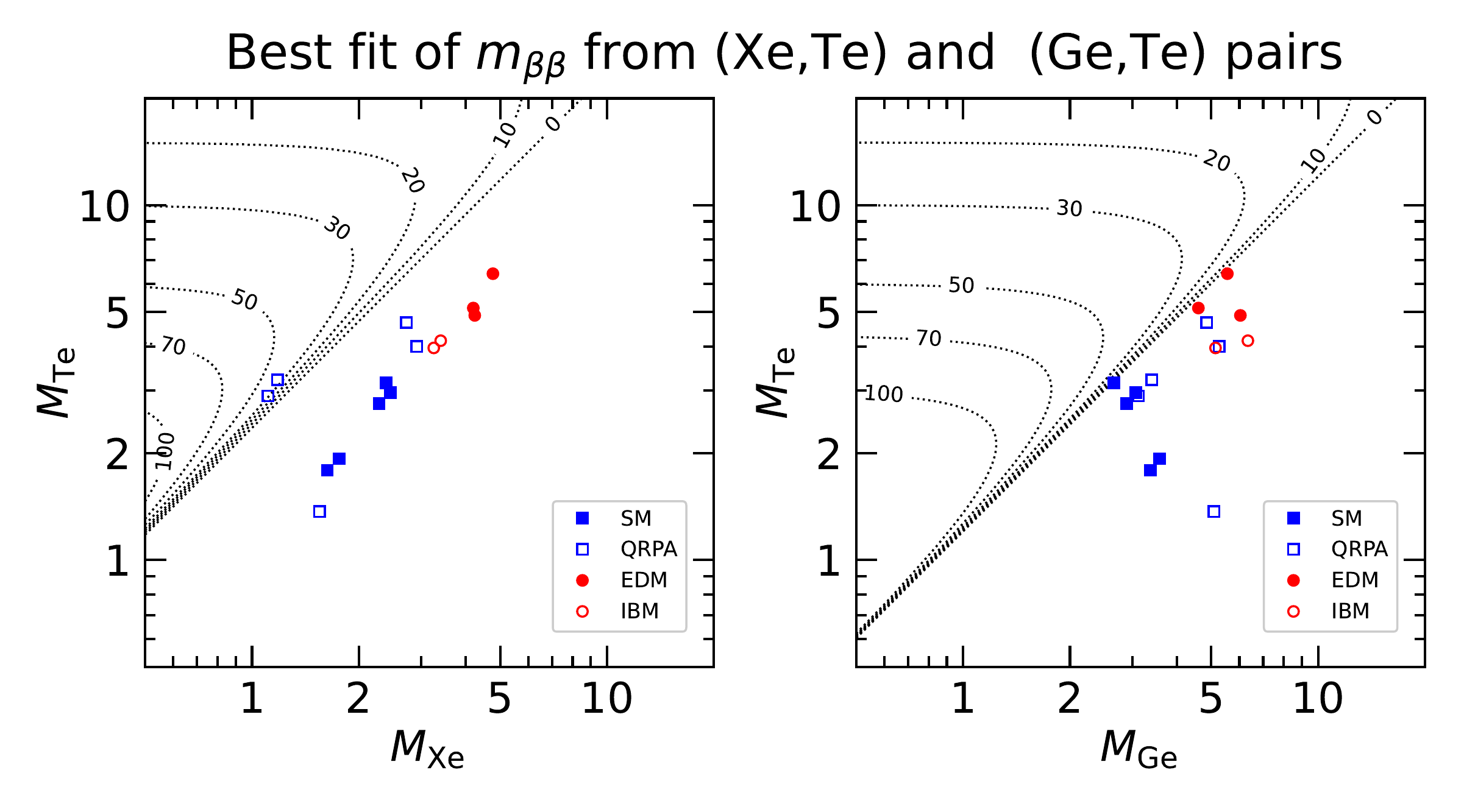}
\vspace*{-3mm}
\caption{\label{Fig_05}
\footnotesize Isolines of best-fit values of $m_{\beta\beta}/\mathrm{meV}$ in the planes charted by (Xe,~Te)
and (Ge,~Te) nuclear matrix elements. The best fit is zero in the lower-right parts of each panel,
as well as in the whole (Xe,~Ge) panel (not shown). 
} \end{minipage}
\end{figure}

Concerning the upper half of Table~\ref{tab:Te}, the 
combination of Te with Xe (or Ge) does not necessarily improve upon the 
separate $2\sigma$ bounds. 
Roughly speaking, when the best-fit value of $m_{\beta\beta}$ in Te is comparable or 
larger than the upper bound from Xe (Ge) alone, the joint bound from Xe+Te (Ge+Te) is weakened, as a result
of the slight tension between the two isotopic data. A slight weakening also occurs whenever when Te is added
to Xe+Ge in the global combination Xe+Ge+Te.%
\footnote{ These effects are analogous to those noted in Sec.~\ref{Subsec:ExpInput}
for the combination of KamLAND-Zen and EXO data,
leading to a $T_{90}$ bound slightly weaker than from KamLAND-Zen alone, as a result of two opposite
fluctuations.}
\textcolor{black}{The $2\sigma$ bounds on $m_{\beta\beta}$ compiled in Table~\ref{tab:Te} are contained in the following range:}
\begin{equation}
\textcolor{black}{m_{\beta\beta} \in   [40.5,\, 135.0] ~\mathrm{meV\ (Xe+Ge+Te) }\ .}
\label{summary} 
\end{equation}
The lowest (most optimistic) edge of this range would significantly cut from above 
the IO and NO allowed regions in Fig.~\ref{Fig_01}, setting also an upper limit on $\Sigma$  
at the level of $\sim 400$~meV.

Figure~\ref{Fig_05} shows isolines of the best-fit value of $m_{\beta\beta}$ in the left and right panels charted by (Xe,~Te)
and (Ge,~Te), respectively. In the left panel, the condition $\sum_i\beta_i>0$ for $m_{\beta\beta}>0$
implies $M_\mathrm{Te}/M_\mathrm{Xe}>2.42$, satisfied only by two marked points (corresponding to the QRPA cases 9 and 10 in Table~\ref{tab:NME}) close to the isoline at $m_{\beta\beta}= 30$~meV.
 In the right panel, none of the marked points 
falls in the analogous region
 $M_\mathrm{Te}/M_\mathrm{Ge}>1.21$, although three of them are very close to its border. 
If future $0\nu\beta\beta$ experiments will show some indications for a signal, plots
like these will help to locate the best-fit values of $m_{\beta\beta}$ as a function of the NME,
for each isotope pair. The (in)consistency of the best fits in different pairs will provide interesting 
clues about the interpretation of data in terms of light Majorana neutrinos.

\begin{figure}[b!]
\vspace*{-2mm}
\begin{minipage}[c]{0.85\textwidth}
\includegraphics[width=0.73\textwidth]{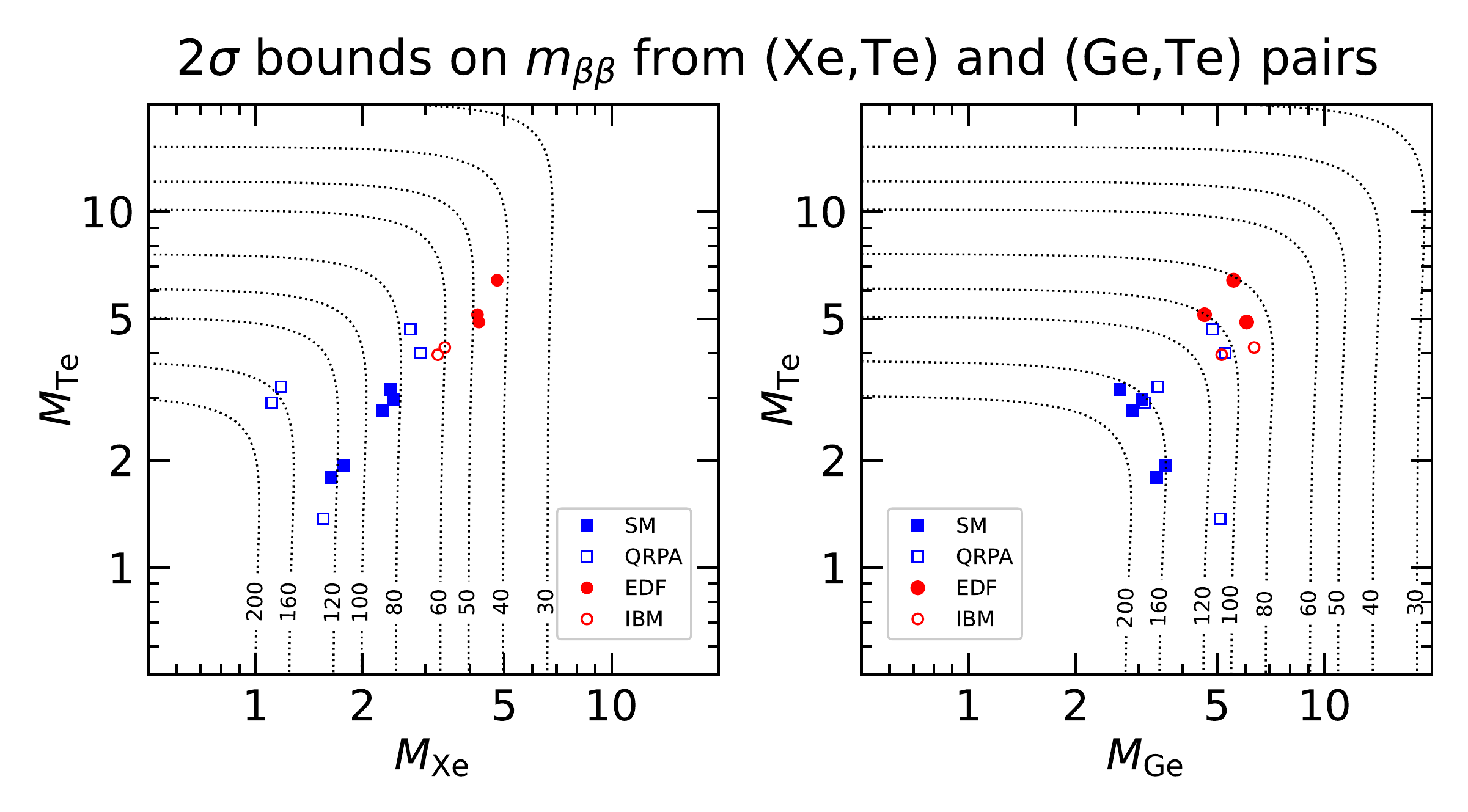}
\vspace*{-3mm}
\caption{\label{Fig_06}
\footnotesize Isolines of $m_{\beta\beta}$ bounds (at $2\sigma$ level)  in the planes charted by (Xe,~Te)
and (Ge,~Te) nuclear matrix elements. The (Xe,~Ge) panel (not shown) is unchanged with respect to Fig.~\ref{Fig_04}.
} \end{minipage}
\end{figure}

Figure~\ref{Fig_06} shows isolines of the $m_{\beta\beta}$ upper bounds at $2\sigma$, in the same planes 
of Fig.~\ref{Fig_05}. In comparison with the lower
panels of Fig.~\ref{Fig_04}, a slight weakening of the bounds can be appreciated. Note that,
in the  presence of subregions where $m_{\beta\beta}>0$ at best fit, the offset $\gamma$ depends on information
coming from both isotopes, whose $\chi^2$ contributions cannot be separated in the combination. 
The condition of equal contributions to $\Delta_\mathrm{CL}$ cannot be defined in general terms, and 
the dashed lines of Fig.~\ref{Fig_04} are thus absent in Fig.~\ref{Fig_06}.

We conclude this Section by discussing the constraints on $m_{\beta\beta}$ at various confidence levels,
as derived from the global combination of current (Xe+Ge+Te) data, using the representative NME values in Table~\ref{tab:NME}.
Since the best fit is $m_{\beta\beta}=0$ in all Xe+Ge+Te cases (see Tab.~\ref{tab:Te}), only upper bounds need to be quoted. 

\begin{figure}[b!]
\begin{minipage}[c]{0.85\textwidth}
\includegraphics[width=0.8\textwidth]{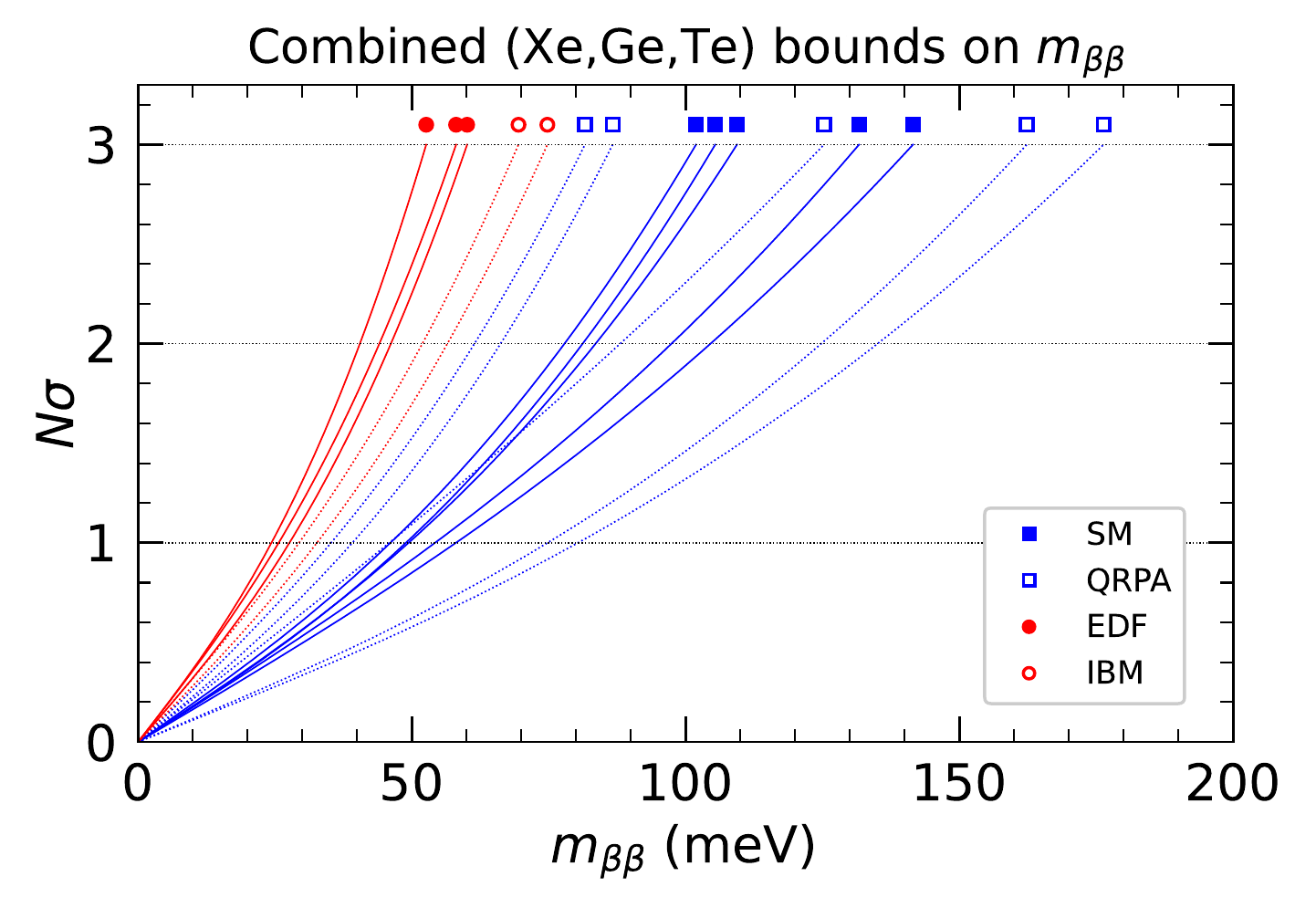}
\vspace*{-3mm}
\caption{\label{Fig_07}
\footnotesize Significance of upper limits on $m_{\beta\beta}$ in terms of $N_\sigma=(\Delta\chi^2)^{1/2}$, from
the combination of current Xe+Ge+Te data, for the representative NME calculations considered in this work. 
} \end{minipage}
\end{figure}

Table~\ref{tab:3sigma} reports the 90\% C.L., $2\sigma$, and $3\sigma$ upper limits on $m_{\beta\beta}$ (in meV). 
Figure~\ref{Fig_07} shows the $N_\sigma=(\Delta\chi^2)^{1/2}$ bounds as continuous functions of
$m_{\beta\beta}$. Qualitatively, the strongest limits in  
are obtained using NMEs from the EDF and IBM models, followed by QRPA and SM cases in mixed order. 
These results can be generalized to any other choice of NME calculations, using the information
provided in this paper.

\begin{table}[t!]
\centering
\resizebox{.4\textwidth}{!}
{\begin{minipage}{0.45\textwidth}
\caption{\label{tab:3sigma} 
Upper bounds on $m_{\beta\beta}/\mathrm{meV}$ at 90\% C.L., $2\sigma$, and $3\sigma$, from the 
combination of current Xe+Ge+Te data, for the representative NME calculations considered in this work. }
\begin{ruledtabular}
\begin{tabular}{rrrrc}
 & 90\% & $2\sigma$ & $3\sigma$ & Model \\ 
\hline
1 &  72.7 &  83.5 & 109.4 &     \\
2 &  67.8 &  77.9 & 101.9 &     \\
3 &  89.3 & 104.5 & 141.5 &  SM \\
4 &  83.3 &  97.4 & 131.7 &     \\
5 &  71.0 &  81.2 & 105.4 &     \\
\hline
6 &  73.6 &  88.0 & 125.3 &     \\
7 &  53.0 &  61.4 &  81.6 &     \\
8 &  57.6 &  66.2 &  86.7 & QRPA\\
9 & 117.7 & 135.0 & 176.3 &     \\
10& 108.9 & 124.7 & 162.3 &     \\
\hline
11&  40.3 &  46.2 &  60.1 &     \\
12&  35.4 &  40.5 &  52.6 & EDF \\
13&  38.2 &  44.1 &  58.1 &     \\
\hline
14&  48.8 &  56.5 &  74.7 &     \\
15&  44.6 &  51.9 &  69.5 & IBM     
\end{tabular}
\end{ruledtabular}
\end{minipage}}
\end{table}

In Fig.~\ref{Fig_07}, one of the $N_\sigma$ curves labelled as QRPA
shows a a markedly different (almost linear) slope, intersecting three SM curves.
This peculiar curve corresponds to the lowest QRPA point in both panels of Fig.~\ref{Fig_06} 
(case 6 in Table~\ref{tab:NME}), that is characterized 
by a rather large value of $M_\mathrm{Ge}$ as compared with $M_\mathrm{Xe,Te}$. As a result,
Ge data prevail over Xe and Te in the combination, and the leading dependence is
$\Delta\chi^2\propto S \propto m^2_{\beta\beta}$ (rather than 
$\Delta\chi^2\propto S^2 \propto m^4_{\beta\beta}$), implying a roughly linear $N_\sigma(m_{\beta\beta})$ function.
Once more, this observation shows the 
importance of considering  the full likelihood profile of the experimental results (e.g., in terms of $S=1/T$), 
rather than pointlike information (such as the 90\% C.L.\ limit on $T$).

As a final step, one could include a joint probability distribution or a
$\Delta\chi^2$ penalty defined over the NME landscape $(M_\mathrm{Xe},\,M_\mathrm{Ge},\,M_\mathrm{Te})$, and
numerically  minimize the 
total $\Delta\chi^2$ function. 
This exercise was performed in \cite{Capozzi:2021fjo} by assuming a conservative characterization
of the NME and their correlated uncertainties, derived within QRPA calculations \cite{Faessler:2008xj}. 
A limit $m_{\beta\beta}<110$~meV was obtained at $2\sigma$ \cite{Capozzi:2021fjo}.
By repeating the same exercise with the updated (Xe+Ge+Te) combination considered herein, we get the following 
marginalized bounds
(in meV):
$m_{\beta\beta}<79.5$ at 90\% C.L., $m_{\beta\beta}<99.8$ at $2\sigma$, and $m_{\beta\beta}<169$ at $3\sigma$. 
Roughly speaking, from these results and from the summary in Eq.~(\ref{summary}) one can state that
the combination of current $0\nu\beta\beta$ experiments sets $2\sigma$ upper bounds 
on $m_{\beta\beta}$ at the level of $\sim 90$--100~meV
for ``average'' NME values, possibly lowered to $\sim 40$--50~meV for favorable NME values.

A final remark is in order. In principle, one should replace the QRPA input from  \cite{Faessler:2008xj} with more
general and up-to-date estimates of the NME's and their uncertainties, characterizing also
the spread among different models and calculations. 
However, no consensus estimates exist yet for NME fiducial values and covariances, 
although relevant work is in progress toward this goal
\cite{Agostini:2022zub,Engel:2016xgb}. Part of the planned strategy involves benchmarking nuclear models for $0\nu\beta\beta$ 
decay against a variety of data, coming from related electroweak and strong interaction processes or from 
\textcolor{black}{nuclear structure  
\cite{Ejiri:2019ezh,Cirigliano:2022oqy,Horoi:2022ley,Ejiri:2022zdg}.
}

\section{Conclusions and perspectives}
\label{Sec:Sum}

We have discussed an approach to the analysis of neutrinoless double beta decay experiments, in terms
of $\Delta\chi^2$ profiles for the signal strength $S_i$ (inverse of the half-life $T_i$) in
the isotopes $i=$~Xe, Ge and Te, building upon previous work \cite{Capozzi:2021fjo}. 
The approach becomes exceedingly simple for quadratic approximations
to such profiles, implying quadratic (in)equalities in the landscape of nuclear matrix elements $M_i$
that connect the $S_i$ to the Majorana mass $m_{\beta\beta}$. For convenience,
some results have been discussed in terms of pseudo data  for null signal in Te (dubbed Te$^*$).
Simple relations among the $M_i$ have been derived to gauge
the relative contributions of different isotopic data in setting upper limits to $m_{\beta\beta}$ 
(for null best fits in Xe, Ge and Te$^*$), and 
to identify the conditions leading to a preference for nonzero $m_{\beta\beta}$ (for generic Xe, Ge and Te data). 
Using the latest available $0\nu\beta\beta$ data, as well as
representative values of the NME from different models, we have discussed 
current constraints on $m_{\beta\beta}$ at several confidence levels and in various
combinations, both numerically and graphically. Global $2\sigma$ upper limits on $m_{\beta\beta}$ are found in the range from 
40.5 to 135~meV, depending on the NME.

The approach can be easily extended to nonstandard processes for $0\nu\beta\beta$ decay 
\cite{Agostini:2022zub,Dolinski:2019nrj,Rodejohann:2011mu} by replacing the relation $S_i=G_i M^2_i m^2_{\beta\beta}$
with the appropriate phase space, NME and particle physics parameter
characterizing the process. Also, the approach can be extended to generic $\Delta\chi^2(S_i)$ functions,
with a modest price to pay in terms of numerical (rather than analytical) solutions. 
We invite the experimental collaborations involved in $0\nu\beta\beta$ decay searches to publicly provide 
such $\Delta\chi^2(S_i)$ 
functions or equivalent ones, as they contain much more information than the usually quoted 90\% C.L.\ limits on $T_i$.
Indeed, the relative impact of such limits and of the resulting bounds on $m_{\beta\beta}$ in
a multi-isotope combination depend
sensitively on the likelihood profiles of $S_i$ (or $T_i)$, and not only on the
relative size of the $M_i$. 

Our approach to the multi-isotope data analysis
would be complete if one could also assign joint probability densities 
to the $M_i$, whose variations could then be treated as nuisance parameters and marginalized. 
So far, detailed results for the NME central values and covariances,
including $g_A$ quenching uncertainties, have been obtained in a specific (QRPA) model \cite{Faessler:2008xj}. In perspective,
it would be important to extend such investigations to other nuclear models, eventually reaching
consensus values for the  $M_i$ and for their correlated (and possibly reduced) uncertainties. 

In this sense, the combined analysis of $0\nu\beta\beta$ results
is proceeding through to the same steps 
that have characterized similar fields (e.g., solar neutrinos) in the past: from low-statistics data
and theoretical models with large uncertainties, to a wealth 
of accurate experimental results interpreted in increasingly refined and constrained models.   
Our work aims at providing one methodological step along this path.

\newpage

\acknowledgments
  
This work is partly supported by the Italian Ministero dell'Universit\`a e Ricerca (MUR) through
the research grant number 2017W4HA7S ``NAT-NET: Neutrino and Astroparticle Theory Network'' under the program PRIN 2017,
and by the Istituto Nazionale di Fisica 
Nucleare (INFN) through the ``Theoretical Astroparticle Physics''  (TAsP) project. We thank K.\ Inoue for 
useful information about the latest KamLAND-Zen data release \cite{Private}.



\vspace*{2mm}

\begin{thebibliography}{99}

\bibitem{Agostini:2022zub}
M.~Agostini, G.~Benato, J.~A.~Detwiler, J.~Men\'endez and F.~Vissani,
``Toward the discovery of matter creation with neutrinoless double-beta decay,''
[arXiv:2202.01787 [hep-ex]].


\bibitem{Zyla:2020zbs}
P.~A.~Zyla \textit{et al.} [Particle Data Group],
``Review of Particle Physics,''
Prog.\ Theor.\ Exp.\ Phys.\ \textbf{2020}, no.8, 083C01 (2020).

\bibitem{Capozzi:2021fjo}
F.~Capozzi, E.~Di Valentino, E.~Lisi, A.~Marrone, A.~Melchiorri and A.~Palazzo,
``Unfinished fabric of the three neutrino paradigm,''
Phys. Rev. D \textbf{104}, no.8, 083031 (2021)
[arXiv:2107.00532 [hep-ph]].

\bibitem{Abazajian:2022ofy}
K.~N.~Abazajian, N.~Blinov, T.~Brinckmann, M.~C.~Chen, Z.~Djurcic, P.~Du, M.~Escudero, M.~Gerbino, E.~Grohs and S.~Hagstotz, \textit{et al.} ``Synergy between cosmological and laboratory searches in neutrino physics: a white paper,''
[arXiv:2203.07377 [hep-ph]].

\bibitem{KamLAND-Zen:2022tow}
S.~Abe \textit{et al.} [KamLAND-Zen],
``First Search for the Majorana Nature of Neutrinos in the Inverted Mass Ordering Region with KamLAND-Zen,''
[arXiv:2203.02139 [hep-ex]].

\bibitem{Anton:2019wmi}
G.~Anton \textit{et al.} [EXO-200],
``Search for Neutrinoless Double-$\beta$ Decay with the Complete EXO-200 Dataset,''
Phys. Rev. Lett. \textbf{123}, no.16, 161802 (2019)
[arXiv:1906.02723 [hep-ex]].

\bibitem{Agostini:2020xta}
M.~Agostini \textit{et al.} [GERDA],
``Final Results of GERDA on the Search for Neutrinoless Double-$\beta$ Decay,''
Phys. Rev. Lett. \textbf{125}, no.25, 252502 (2020)
[arXiv:2009.06079 [nucl-ex]].

\bibitem{Alvis:2019sil}
S.~I.~Alvis \textit{et al.} [MAJORANA],
``A Search for Neutrinoless Double-Beta Decay in $^{76}$Ge with 26 kg-yr of Exposure from the MAJORANA DEMONSTRATOR,''
Phys. Rev. C \textbf{100}, no.2, 025501 (2019)
[arXiv:1902.02299 [nucl-ex]].

\bibitem{Adams:2021rbc}
D.~Q.~Adams \textit{et al.} [CUORE],
``Search for Majorana neutrinos exploiting millikelvin cryogenics with CUORE,''
 Nature {\bf 604}, 53 (2022); see also:
``High sensitivity neutrinoless double-beta decay search with one tonne-year of CUORE data,''
[arXiv:2104.06906 [nucl-ex]]. 

\bibitem{Caldwell:2017mqu}
A.~Caldwell, A.~Merle, O.~Schulz and M.~Totzauer,
``Global Bayesian analysis of neutrino mass data,''
Phys. Rev. D \textbf{96}, no.7, 073001 (2017)
[arXiv:1705.01945 [hep-ph]].

\bibitem{Biller:2021bqx}
S.~D.~Biller,
``Combined constraints on Majorana masses from neutrinoless double beta decay experiments,''
Phys. Rev. D \textbf{104}, no.1, 012002 (2021)
[arXiv:2103.06036 [hep-ex]].


\bibitem{Markisch:2018ndu}
B.~M\"arkisch, H.~Mest, H.~Saul, X.~Wang, H.~Abele, D.~Dubbers, M.~Klopf, A.~Petoukhov, C.~Roick and T.~Soldner, \textit{et al.}
``Measurement of the Weak Axial-Vector Coupling Constant in the Decay of Free Neutrons Using a Pulsed Cold Neutron Beam,''
Phys. Rev. Lett. \textbf{122}, no.24, 242501 (2019)
[arXiv:1812.04666 [nucl-ex]].


\bibitem{Suhonen:2017krv}
J.~T.~Suhonen,
``Value of the Axial-Vector Coupling Strength in \ensuremath{\beta} and \ensuremath{\beta}\ensuremath{\beta} Decays: A Review,''
Front. in Phys. \textbf{5}, 55 (2017)
[arXiv:1712.01565 [nucl-th]].

\bibitem{Private}
K.\ Inoue, private communication. We thank the KamLAND Collaboration
for sharing the digitized likelihood function used in \cite{KamLAND-Zen:2022tow}. 


\bibitem{Menendez:2017fdf}
J.~Men\'endez,
``Neutrinoless $\beta\beta$ decay mediated by the exchange of light and heavy neutrinos: The role of nuclear structure correlations,''
J. Phys. G \textbf{45}, no.1, 014003 (2018)
[arXiv:1804.02105 [nucl-th]].

\bibitem{Horoi:2015tkc}
M.~Horoi and A.~Neacsu,
``Shell model predictions for $^{124}$Sn double-$\beta$ decay,''
Phys. Rev. C \textbf{93}, no.2, 024308 (2016)
[arXiv:1511.03711 [nucl-th]].

\bibitem{Coraggio:2020hwx}
L.~Coraggio, A.~Gargano, N.~Itaco, R.~Mancino and F.~Nowacki,
``Calculation of the neutrinoless double-$\beta$ decay matrix element within the realistic shell model,''
Phys. Rev. C \textbf{101}, no.4, 044315 (2020)
[arXiv:2001.00890 [nucl-th]].

\bibitem{Mustonen:2013zu}
M.~T.~Mustonen and J.~Engel,
``Large-scale calculations of the double-\ensuremath{\beta} decay of $^{76}$Ge, $^{130}$Te, $^{136}$Xe, and $^{150}$Nd in the deformed self-consistent Skyrme quasiparticle random-phase approximation,''
Phys. Rev. C \textbf{87}, no.6, 064302 (2013)
doi:10.1103/PhysRevC.87.064302
[arXiv:1301.6997 [nucl-th]].

\bibitem{Hyvarinen:2015bda}
J.~Hyv\"arinen and J.~Suhonen,
``Nuclear matrix elements for $0\nu\beta\beta$ decays with light or heavy Majorana-neutrino exchange,''
Phys. Rev. C \textbf{91}, no.2, 024613 (2015)

\bibitem{Simkovic:2018hiq}
F.~\v{S}imkovic, A.~Smetana and P.~Vogel,
``$0\nu\beta\beta$ nuclear matrix elements, neutrino potentials and $\mathrm{SU}(4)$ symmetry,''
Phys. Rev. C \textbf{98}, no.6, 064325 (2018)
[arXiv:1808.05016 [nucl-th]].

\bibitem{Fang:2018tui}
D.~L.~Fang, A.~Faessler and F.~Simkovic,
``0\ensuremath{\nu}$\beta\beta$ -decay nuclear matrix element for light and heavy neutrino mass mechanisms from deformed quasiparticle random-phase approximation calculations for $^{76}$Ge, $^{82}$Se, $^{130}$Te, $^{136}$Xe , and $^{150}$Nd with isospin restoration,''
Phys. Rev. C \textbf{97}, no.4, 045503 (2018)
[arXiv:1803.09195 [nucl-th]].

\bibitem{Rodriguez:2010mn}
T.~R.~Rodriguez and G.~Martinez-Pinedo,
``Energy density functional study of nuclear matrix elements for neutrinoless $\beta\beta$ decay,''
Phys. Rev. Lett. \textbf{105}, 252503 (2010)
[arXiv:1008.5260 [nucl-th]].

\bibitem{LopezVaquero:2013yji}
N.~L\'opez Vaquero, T.~R.~Rodr\'\i{}guez and J.~L.~Egido,
``Shape and pairing fluctuations effects on neutrinoless double beta decay nuclear matrix elements,''
Phys. Rev. Lett. \textbf{111}, no.14, 142501 (2013)
[arXiv:1401.0650 [nucl-th]].

\bibitem{Song:2017ktj}
L.~S.~Song, J.~M.~Yao, P.~Ring and J.~Meng,
``Nuclear matrix element of neutrinoless double-$\beta$ decay: Relativity and short-range correlations,''
Phys. Rev. C \textbf{95}, no.2, 024305 (2017)
[arXiv:1702.02448 [nucl-th]].

\bibitem{Barea:2015kwa}
J.~Barea, J.~Kotila and F.~Iachello,
``$0\nu\beta\beta$ and $2\nu\beta\beta$ nuclear matrix elements in the interacting boson model with isospin restoration,''
Phys. Rev. C \textbf{91}, no.3, 034304 (2015)
[arXiv:1506.08530 [nucl-th]].

\bibitem{Deppisch:2020ztt}
F.~F.~Deppisch, L.~Graf, F.~Iachello and J.~Kotila,
``Analysis of light neutrino exchange and short-range mechanisms in $0\nu\beta\beta$ decay,''
Phys. Rev. D \textbf{102}, no.9, 095016 (2020)
[arXiv:2009.10119 [hep-ph]].



\bibitem{Faessler:2008xj}
A.~Faessler, G.~L.~Fogli, E.~Lisi, V.~Rodin, A.~M.~Rotunno and F.~Simkovic,
``QRPA uncertainties and their correlations in the analysis of $0\nu\beta\beta$ decay,''
Phys. Rev. D \textbf{79}, 053001 (2009)

\bibitem{Stoica:2019ajg}
\textcolor{black}{
S.~Stoica and M.~Mirea,
``Phase Space Factors for Double-Beta Decays,''
Front. in Phys. \textbf{7}, 12 (2019)
}

\bibitem{Bilenky:2002ga}
S.~M.~Bilenky and J.~A.~Grifols,
``The Possible test of the calculations of nuclear matrix elements of the $(\beta\beta)_{0\nu}$ decay,''
Phys. Lett. B \textbf{550}, 154-159 (2002)
[arXiv:hep-ph/0211101 [hep-ph]].

\bibitem{Bilenky:2004um}
S.~M.~Bilenky and S.~T.~Petcov,
``Nuclear matrix elements of $0 \nu \beta \beta$ decay: Possible test of the calculations,''
[arXiv:hep-ph/0405237 [hep-ph]].

\bibitem{Engel:2016xgb}
J.~Engel and J.~Men\'endez,
``Status and Future of Nuclear Matrix Elements for Neutrinoless Double-Beta Decay: A Review,''
Rept. Prog. Phys. \textbf{80}, no.4, 046301 (2017)
[arXiv:1610.06548 [nucl-th]].

\bibitem{Ejiri:2019ezh}
H.~Ejiri, J.~Suhonen and K.~Zuber,
``Neutrino\textendash{}nuclear responses for astro-neutrinos, single beta decays and double beta decays,''
Phys. Rept. \textbf{797}, 1-102 (2019)


\bibitem{Cirigliano:2022oqy}
V.~Cirigliano, Z.~Davoudi, W.~Dekens, J.~de Vries, J.~Engel, X.~Feng, J.~Gehrlein, M.~L.~Graesser, L.~Gr\'af and H.~Hergert, \textit{et al.}
``Neutrinoless Double-Beta Decay: A Roadmap for Matching Theory to Experiment,''
[arXiv:2203.12169 [hep-ph]].

\bibitem{Horoi:2022ley}
M.~Horoi, A.~Neacsu and S.~Stoica,
``A Statistical Analysis for the Neutrinoless Double-Beta Decay Matrix element of $^{48}$Ca,''
[arXiv:2203.10577 [nucl-th]].

\bibitem{Ejiri:2022zdg}
\textcolor{black}{H.~Ejiri, L.~Jokiniemi and J.~Suhonen,
``Nuclear matrix elements for neutrinoless $\beta \beta$ decays and spin-dipole giant resonances,''
Phys. Rev. C \textbf{105}, no.2, L022501 (2022)
[arXiv:2202.00361 [nucl-th]].
}


\bibitem{Dolinski:2019nrj}
M.~J.~Dolinski, A.~W.~P.~Poon and W.~Rodejohann,
``Neutrinoless Double-Beta Decay: Status and Prospects,''
Ann. Rev. Nucl. Part. Sci. \textbf{69}, 219-251 (2019)
[arXiv:1902.04097 [nucl-ex]].

\bibitem{Rodejohann:2011mu}
W.~Rodejohann,
``Neutrino-less Double Beta Decay and Particle Physics,''
Int. J. Mod. Phys. E \textbf{20}, 1833-1930 (2011)
[arXiv:1106.1334 [hep-ph]].




\end{thebibliography}
\end{document}